\documentclass[pdflatex,sn-mathphys-num,iicol]{sn-jnl}

\usepackage{graphicx}%
\usepackage{multirow}%
\usepackage{amsmath,amssymb,amsfonts}%
\usepackage{amsthm}%
\usepackage{mathrsfs}%
\usepackage[title]{appendix}%
\usepackage{xcolor}%
\usepackage{textcomp}%
\usepackage{manyfoot}%
\usepackage{booktabs}%
\usepackage{algorithm}%
\usepackage{algorithmicx}%
\usepackage{algpseudocode}%
\usepackage{listings}%
\usepackage{xspace}
\usepackage{import}
\usepackage{float}
\usepackage{xurl}

\theoremstyle{thmstyleone}%
\theoremstyle{thmstyletwo}%

\theoremstyle{thmstylethree}%

\unnumbered

\newcommand{\water}{H\textsubscript{2}O\xspace}
\newcommand{\coso}{\texttt{TRECI}}
\newcommand{\cosot}{\coso~}
\newcommand{\Franken}{\texttt{Franken}}
\newcommand{\Frankent}{\Franken~}
\newcommand{\MACE}{\texttt{MACE}}
\newcommand{\MACEt}{\MACE~}
\newcommand{\FrankenMP}{\texttt{Franken-MP0}}
\newcommand{\FrankenMPt}{\FrankenMP~}
\newcommand{\FrankenPZC}{\texttt{Franken-PZC}}
\newcommand{\FrankenPZCt}{\FrankenPZC~}
\newcommand{\DEAL}{\texttt{DEAL}}
\newcommand{\DEALt}{\DEAL~}
\newcommand{\subsub}[1]{\vspace{0.15cm}\textbf{#1}.}

\begin{document}
\title{Electrified Metal-Water Interfaces at Constant Potential: Data-Efficient Transfer Learning for Machine-Learning-Based Molecular Dynamics}  

\author[1,2]{\fnm{Michele Giovanni} \sur{Bianchi}}
\author[1]{\fnm{Michele} \sur{Re Fiorentin}}
\author[1]{\fnm{Francesca} \sur{Risplendi}}
\author[1,3]{\fnm{Candido Fabrizio} \sur{Pirri}}
\author[2]{\fnm{Michele} \sur{Parrinello}}
\author*[2]{\fnm{Luigi} \sur{Bonati}}\email{luigi.bonati@iit.it}
\author*[1]{\fnm{Giancarlo} \sur{Cicero}}\email{giancarlo.cicero@polito.it}

\affil[1]{\orgdiv{Department of Applied Science and Technology}, \orgname{Politecnico di Torino}, \orgaddress{\street{corso Duca degli Abruzzi 24}, \city{Torino}, \postcode{10129}, \state{Italy}}}

\affil[2]{\orgdiv{Atomistic Simulations}, \orgname{Italian Institute of Technology}, \orgaddress{\street{via Enrico Melen 83}, \city{Genova}, \postcode{16152}, \state{Italy}}}

\affil[3]{\orgdiv{Centre for
Sustainable Future Technologies}, \orgname{Italian Institute of Technology}, \orgaddress{\street{via Livorno 60}, \city{Torino}, \postcode{10144}, \state{Italy}}}


\abstract{Simulating electrified metal-water interfaces with explicit solvent under constant potential is essential for understanding electrochemical processes, yet remains prohibitively expensive with \textit{ab initio} methods. We present \coso, a data-efficient workflow for constructing machine learning force-fields (ML-FFs) that achieve \textit{ab initio}-level accuracy in electronically grand-canonical molecular dynamics. By leveraging transfer learning from general-purpose and domain-specific models, \cosot enables stable and accurate simulations across a wide potential range using a reduced number of reference configurations. This efficiency allows the use of high-level meta-GGA functionals and rigorous surface-electrification schemes. Applied to Cu(111)/water, models trained on just one thousand configurations yield accurate molecular dynamics simulations, capturing bias-dependent solvent restructuring effects not previously reported. \cosot offers a general strategy for characterising diverse materials and interfacial chemistries, significantly lowering the cost of realistic constant-potential simulations and expanding access to quantitative electrochemical modelling.}




\maketitle

\section{Introduction}\label{intro}
Charged metal-\water interfaces play an important role in fields such as corrosion, heterogeneous catalysis, and electrochemistry~\cite{carrasco_molecular_2012,gonella_water_2021}. Despite decades of study, recently intensified by the rise of electrochemical technologies for the energy transition, understanding these interfaces at the atomic level remains a major challenge in surface science~\cite{kolb_reconstruction_1996,xia_electric-field_1995,iwasita_situ_1997,cicero_anomalous_2011,nitopi_progress_2019,auer_potential_2021,zhu_machine_2025}. \textit{Ab initio} molecular dynamics (AIMD) simulations are an ideal candidate to provide atomistic insight in these systems~\cite{gros_ab_2022}. However, the heterogeneity of the interfaces composed of metal surfaces, solvents and solutes clashes with the problem of small spatial and short time scales achievable in AIMD. In recent years, machine learning interatomic potentials, here referred to as ML force-fields (ML-FFs) to avoid confusion with the applied electric potential, have emerged as a promising solution to overcome the limits of AIMD. They bridge the accuracy of \textit{ab initio} approaches with the computational feasibility of force-field schemes~\cite{unke_machine_2021}. However, the presence of an applied electrical bias significantly increases the simulation complexity both at the \textit{ab initio} and ML level. The simplest approaches for the electrification of the interface are based on constant charge schemes~\cite{le_recent_2021}, which are not fully representative of the experiments, since the applied bias is not controllable. In principle, constant potential methods align more closely with experimental ensembles~\cite{hormann_converging_2023}. However, they are more computationally expensive, and in practice they are applied in conjunction with implicit solvent schemes~\cite{kastlunger_controlled-potential_2018,islam_implicit_2023,le_explicit-implicit_2023} which can introduce artefacts. In addition to these issues associated with the electrification schemes, ML-FFs face a further complication: the treatment of long-range electrostatic interactions. The presence of charged species can give rise to interactions that extend beyond the local nature of the force-fields~\cite{ko_fourth-generation_2021,zhang_molecular-scale_2024}.  
As a result, the development of ML-FFs for electrochemical systems constitutes an area of research that remains largely unexplored. Preliminary ML-FFs in the electrochemistry domain neglect the effects of the applied bias, focusing only on the potential of zero charge (PZC), i.e., when there is no extra charge on the surface~\cite{natarajan_RCS_2016,mikkelsen_is_2021,rice_hydrogen_2021}. To overcome these limitations, ML-FFs employing a constant charge approach were developed~\cite{zhu_machine_2025}, while subsequent advances proposed constant potential ML models employing hybrid explicit-implicit solvent schemes~\cite{chen_atomistic_2023,bergmann_machine_2025,tian_electrochemical_2025,chen2024constantpotentialreactorframework}.
\begin{figure*}[h]
\centering
\includegraphics[width=0.8\textwidth]{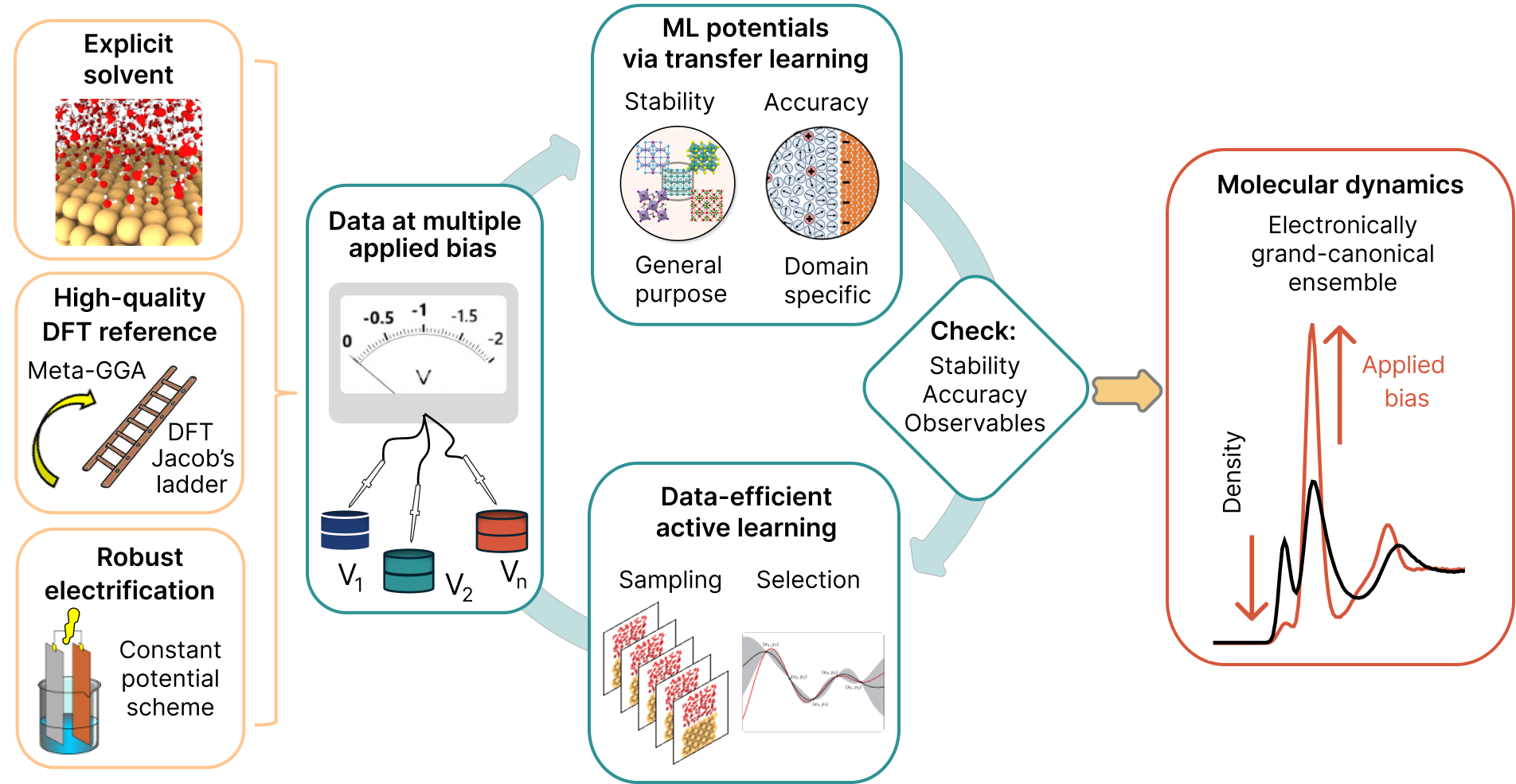}
\caption{General scheme of \cosot for electronically grand-canonical ML-FFs: the pillars of the methods are a high-quality dataset at multiple applied bias values, a data-efficient ML architecture via transfer learning and an effective active learning approach.}
\label{fig:scheme_general}
\end{figure*}

In this work, we present \coso, a data-efficient strategy that uniquely leverages transfer learning~\cite{chen_data-efficient_2023,zaverkin_transfer_2023} and active learning to construct ML-FFs tailored for electrified metal-water interfaces. By featuring a fully explicit solvent model and a rigorous constant potential scheme (the ``double reference method''~\cite{taylor_first_2006}), \cosot allows for higher fidelity simulations, with a level of accuracy that was not possible before.
We apply \cosot to a copper-water interface, a representative electrode used in many electrocatalytic applications~\cite{nitopi_progress_2019,naher_emerging_2025}. \cosot achieves stable and accurate simulations with only a thousand configurations, a significant reduction of the dataset size in comparison to similar models trained from scratch. Notably, \cosot enables direct observation of bias-dependent solvent restructuring phenomena, offering insights that were inaccessible with prior models.
The \cosot code is openly available at~\url{https://github.com/michelegiovannibianchi/TRECI}, and its versatility allows straightforward application to other electrochemical systems. This paves the way for broader access to advanced constant potential simulations in a wide range of electrochemical applications.

\section{Results}\label{results}
\subsection{Data-efficient workflow for electrochemical interfaces}\label{core_element}
The main outcome is the development of \cosot (TRansfer learning for ElectroChemical Interfaces), a computational workflow for ML-FFs that provides a high-fidelity description of electrified metal-water interfaces while remaining data-efficient.
The success of \cosot relies on three core pillars: 1) a high-quality quantum mechanical description of the interface, 2) a data-efficient transfer learning framework for building the ML-FFs and 3) an effective active learning strategy for collecting the required configurations (Fig.~\ref{fig:scheme_general}).
\begin{figure*}[!tbp]
\centering
\includegraphics[width=0.7\textwidth]{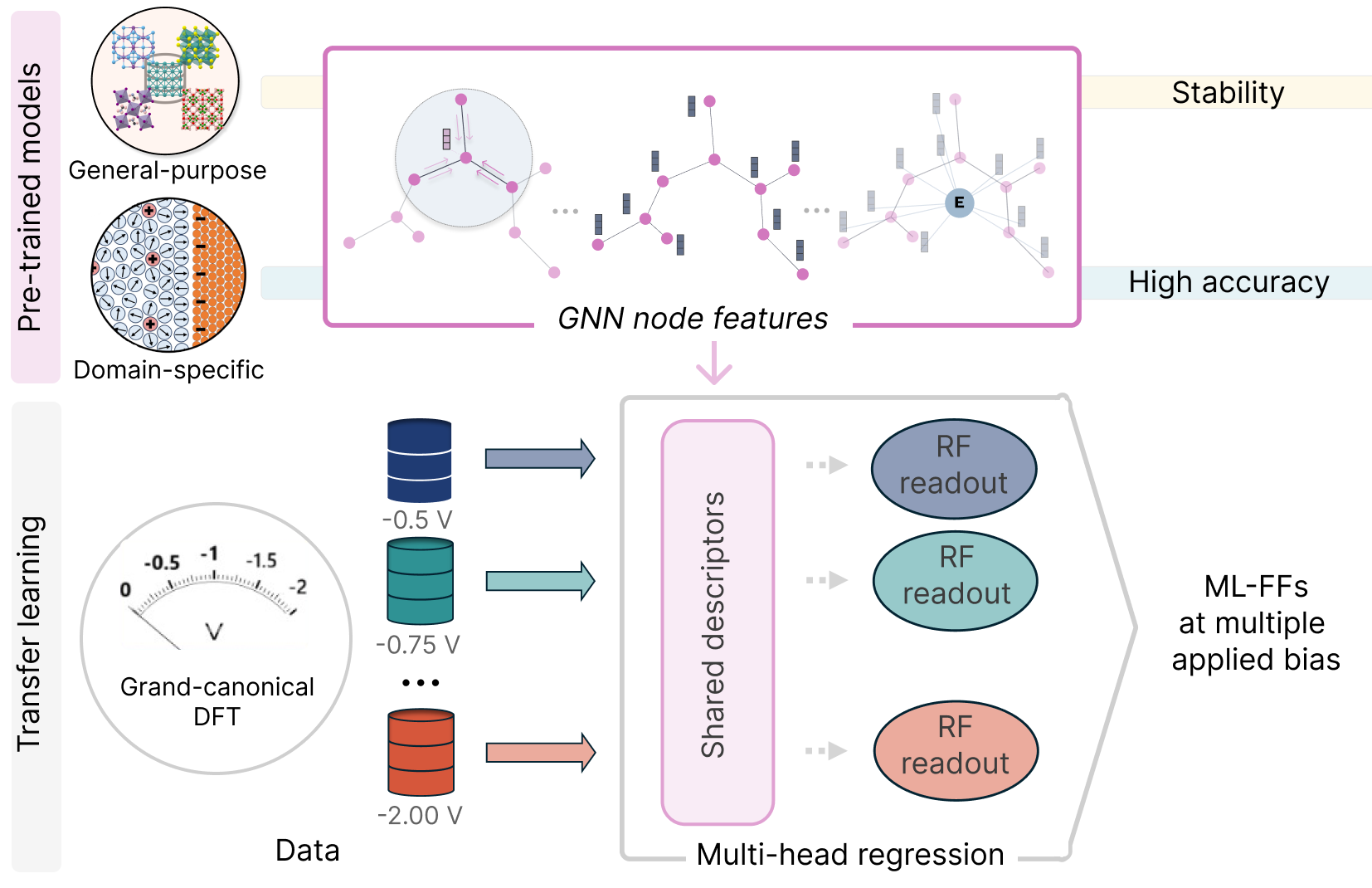}
\caption{\cosot strategy for the training of the constant potential ML-FFs via transfer learning: the node features of a pre-trained general-purpose or domain-specific GNNs are used as descriptors for a multi-head \Frankent model. Each readout block implements a large-scale kernel regression model targeting an applied bias value. Kernel functions are approximated by means of Random Fourier Features (RF)}
\label{fig:scheme_transfer_learning}
\end{figure*}

\subsub{High-quality description at multiple applied bias values}\label{high_data}
A high-quality dataset and rigorous labelling are essential elements to provide an accurate description of electrified interfaces via ML-FFs. Any artifact in the solvent representation or ambiguities in the definition of the electrode potential can introduce inconsistencies in the labels, which become particularly detrimental when working with the small datasets employed in this work.\\
\cosot achieves a high-quality modelling of the interfaces via a) a comprehensive description of the solvent dynamics, b) an accurate treatment of the liquid water network, and c) a robust electrification scheme (cf. left side of Fig.~\ref{fig:scheme_general}).\\
The comprehensive description of the solvent is achieved by following a fully explicit solvent approach without any artificial vacuum or implicit solvent region. This allows for capturing in the dataset both the overall dielectric response of the solvent and the localised interactions between molecules and the surface.\\
A high-quality description of the liquid water network implies the choice of advanced DFT functionals due to the well-known shortcomings of the common PBE-GGA approximation in water-based systems~\cite{Chen_PNAS_2017,gartner_PNAS_2020}. We develop the datasets using the meta-GGA SCAN functional~\cite{sun_PRL_2015} as the DFT reference (cf. Section ``Methods'').\\
The inclusion of the effects of the applied bias in the framework of ML-FFs represents the most crucial element. This is realised within an electronically grand-canonical ensemble using the ``double reference method''~\cite{taylor_first_2006}. By employing this scheme, \cosot constructs multiple datasets corresponding to distinct potential energy surfaces (PESs), each associated with a well-defined and macroscopically accessible quantity, the bias value.
The ``double reference method'' determines the electrode potential by using a series of potential references computed in auxiliary systems. This approach eliminates artifacts from solvent-vacuum interfaces and avoids inconsistencies between the energies computed in the systems with and without extra electrons (see Section ``Methods'').
As a result, it provides a consistent relationship between surface charge and electrode potential, removes ambiguities in data labelling, and guarantees that each PES is uniquely identified.
Such a rigorous electrification scheme represents the main step ahead with respect to other ML works.\\
To the best of our knowledge, a single framework integrating all these three modelling choices has not been realised in prior studies, largely due to the high computational cost associated with each component: SCAN is approximately 2.5 times more expensive than PBE, and a simulation with the ``double reference method'' is up to 10 times more costly than a calculation without external bias. When used together, these methods become prohibitively expensive, calling for highly data-efficient approaches. 

\subsub{ML force-fields via transfer learning}\label{intro_transfer}
Given the high cost of generating such data across multiple potentials, training independent ML-FFs for each applied bias value would be highly inefficient.
\cosot instead adopts a transfer learning paradigm~\cite{falk2023transfer} that reuses information from models trained in the absence of an electric field to efficiently learn the corresponding constant applied bias PESs.\\
Specifically, we draw on the recently introduced \Frankent framework~\cite{novelli_fast_2025}. In this scheme, energies and forces are predicted using a large-scale kernel regression model. Differently from the hand-crafted descriptors of traditional kernel-based ML-FFs, here they are built upon the node features of a pre-trained graph neural network (GNN), such as the general-purpose potentials from the \texttt{MACE-MP} family~\cite{batatia_foundation_2024}, inheriting the representations learnt by the deep neural network (cf. Section ``Methods'').
Within \coso, we extend this idea into a \textit{multi-head} architecture, where a shared set of GNN-based descriptors is reused to simultaneously learn several PESs corresponding to different applied biases. Each head is trained to address a specific bias value (Fig.~\ref{fig:scheme_transfer_learning}).
Because only a small regression block is bias-specific, the overall data requirement is drastically reduced.
Furthermore, \cosot generalises the \Frankent concept beyond fine-tuning of universal ML-FFs: it also leverages domain-specific models trained for the same system at the PZC, in the absence of an applied field. In the following we refer to models constructed using general-purpose descriptors and domain-specific ones as \FrankenMPt and \FrankenPZC, respectively.
As detailed in the Section ``From stability to accuracy'' below, this combination of general-purpose and domain-specific models provides a route from stable exploratory simulations to accurate production runs.

\subsub{Data-efficient active learning}\label{active_learning}
The efficiency of \cosot ultimately hinges on identifying the few thermodynamically relevant configurations at each target bias. Simply re-labelling equilibrium structures obtained in the absence of an electric field is insufficient, as the applied potential can stabilise entirely different interfacial arrangements.
To address this, \cosot adopts an iterative data acquisition strategy based on active learning. The dataset is progressively expanded by exploring the PESs with preliminary ML-FFs and selecting a limited number of new configurations for DFT labelling.\\
Specifically, for the selection, we exploit the capabilities of the ``Data-Efficient Active Learning'' (\DEAL) protocol~\cite{perego_data_2024}, which evaluates the similarity among the local environments of the sampled configurations to identify a reduced set of geometries to label. This filtering step is based on the Bayesian predictive variance of a Sparse Gaussian Process model (see Section ``Methods''). This approach yields a compact and non-redundant dataset while ensuring efficient coverage of the relevant configurational space.

\begin{figure*}[!htbp]
\centering
\includegraphics[width=0.9\textwidth]{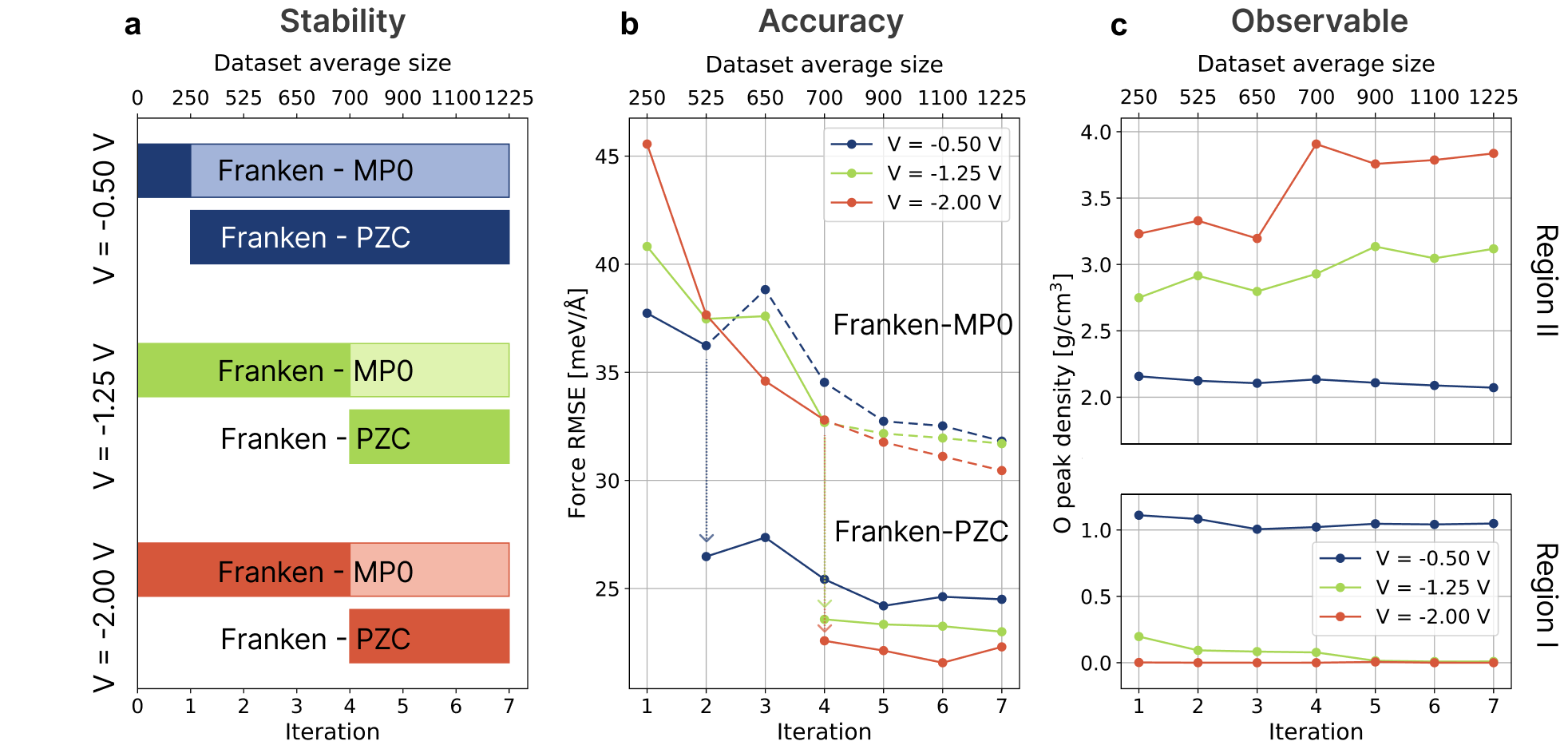}
\caption{Evolution of the learning process for \cosot models at representative target biases during the active learning. (a) Stability range of the different descriptors (\FrankenMPt vs \FrankenPZC): coloured bars show the iterations at which the ML-FFs are stable; dark colours identify the descriptors employed for the sampling at the different iterations. (b) Force RMSE for \FrankenMPt and \FrankenPZC. Solid lines identify the descriptors employed for the sampling. The accuracy is not reported when the models are not stable. (c) Peak value of the oxygen density in the different interfacial regions.}
\label{fig:accuracy_stability}
\end{figure*}
\subsub{Complete strategy}
Here we summarise the steps of \cosot to construct ML-FFs at constant potential. The workflow begins with the construction of an ML-FF trained for the PZC case using the \MACEt architecture (see Section ``Methods''). The PZC model is developed to have domain-specific descriptors useful for the transfer learning. In addition, this gives a starting dataset for the ML-FFs with applied bias and serves as a baseline for the results in the absence of an electric field.\\
An iterative strategy composed of four steps is then applied (Fig.~\ref{fig:scheme_general}). The cycle starts with electronically grand-canonical DFT calculations. Using this data, ML-FFs are optimised for the corresponding constant bias values using transfer learning, employing both general-purpose (from the \texttt{MACE-MP} family) and domain-specific descriptors (from the model trained at the PZC). The models are then benchmarked for stability: if domain-specific descriptors fail to yield stable MD trajectories, the workflow proceeds using the general-purpose ones. Instead, once enough data are available and \FrankenPZCt models become robust, they are preferred due to their superior accuracy. 
Then, new atomic configurations are sampled using MD and selected through a data-efficient active learning protocol. The cycle continues until both the accuracy of energy and force predictions, as well as relevant physical observables, reach convergence. Only at this point, the ML-FFs are considered sufficiently reliable for use in production-level MD simulations.

\subsection{Validation of the workflow}\label{detail_workflow}
We validate the workflow of \cosot on a real case system. In particular, we study the copper-water interface at negative potentials, a representative cathodic system used in electrochemical applications such as the CO\textsubscript{2} electro-reduction reaction~\cite{nitopi_progress_2019}. In Fig.~\ref{fig:accuracy_stability} we report the evolution of the learning process, from which it is clear that after a few active learning iterations and the collection of one thousand structures, \cosot achieves stable, accurate and converged results. 

\subsub{From stability to accuracy}\label{stability_vs_accuracy}
We begin constructing the constant bias models using a subset of 250 configurations selected from the PZC dataset and labelled with grand-canonical DFT. This small dataset proves insufficient for training not only a standard \MACEt architecture but also domain-specific \FrankenPZCt ML-FFs: trajectories generated with these models lack stability.
In contrast, the fine-tuned \FrankenMPt consistently yields stable results across all applied potentials and throughout all active learning cycles (Fig.~\ref{fig:accuracy_stability}a).
For this reason, we start expanding the dataset in an active learning framework using configurations from MD trajectories generated with \FrankenMPt models. As the dataset is enlarged, \FrankenPZCt ML-FFs begin to produce stable results: at low applied bias (\textsf{V}~=~$-0.50$ V), stability is achieved after a single active learning cycle, while more negative potentials require additional iterations (Fig.~\ref{fig:accuracy_stability}a). In particular, we initially use equivariant domain-specific models, which are computationally more expensive but become robust in earlier iterations (Supplementary Section~\ref{SI_Extra_backbone}), and then we transition to invariant ML-FFs for production MD simulations.\\
Switching from general-purpose to domain-specific descriptors leads to a reduction in the prediction error (approximately 30\% lower RMSE), ensuring a uniform accuracy across all the bias values (cf. Fig.~\ref{fig:accuracy_stability}b and Table~\ref{tab:RMSE}). Notably, the accuracy achieved with PZC descriptors is comparable to the corresponding PZC \MACEt model one, but \cosot obtains these results using an amount of data equal to 25\% of the PZC dataset.

\begin{table*}[h]
\centering
\caption{Root-Mean-Squared-Error (RMSE) in meV/\text{\AA} for the prediction of forces using \cosot models at different bias values, with descriptors from either \texttt{MACE-MP0} (\FrankenMP) or our domain-specific PZC model (\FrankenPZC).}
\label{tab:RMSE}
\begin{tabular}{lrrrrrrr}
\toprule
Case [V vs SHE] & $-0.50$ & $-0.75$ & $-1.00$ & $-1.25$ & $-1.50$ & $-1.75$ & $-2.00$ \\
\midrule
\FrankenMP   & 31.8 & 30.9 & 32.2 & 31.7 & 29.8 & 29.9 & 30.5 \\
\FrankenPZC  & 24.5 & 23.7 & 24.3 & 23.0 & 21.3 & 22.1 & 22.3 \\
\bottomrule
\end{tabular}
\end{table*}

\subsub{From accuracy to convergence}
When using \FrankenPZCt models, the force accuracy rapidly converges to a plateau (Fig.~\ref{fig:accuracy_stability}b). Despite this convergence, we perform a more rigorous check by analysing the evolution of key physical observables. As a representative quantity, we monitor the peak heights in the oxygen density profile at the interface, which are highly sensitive to structural changes induced by the applied bias (see Section ``Structure of interfacial water''). 
The analysis of these quantities during the active learning provides insights into the learning process (Fig.~\ref{fig:accuracy_stability}c). As the models are progressively refined, the first peak height (chemisorbed molecules) decreases while the second one (physisorbed water) increases until both converge to stable values. This modulation of the peak density is a well-known effect of the applied potential, and these variations during the active learning cycles are an expected behaviour for a transfer learning process from an electric-field-agnostic model to ML-FFs in the presence of applied bias. 
By tracking these physical observables, we notice that even after the convergence of the force RMSE, approximately three additional active learning iterations are required for the peak densities to reach a plateau.
At this point, we further validate the convergence of the solvent density profiles, comparing \cosot models built upon different descriptors, as detailed in Supplementary Section~\ref{SI_different_backbones}. Even if the starting models for the transfer learning are different, the ML-FFs predict reciprocally consistent observables, confirming the effectiveness of \coso's strategy.

\subsection{Application to Cu(111)-water interface}\label{analysis_interface}
Having validated \coso's workflow, the ML-FFs are employed in large-scale MD simulation to investigate the properties of interfacial water in contact with a Cu(111) surface, focusing on the effects of reducing bias. Based on an estimation of the PZC for this system ($\approx-0.1$~to~$-0.2$~V vs SHE, see Section ``Methods''), we consider values smaller than $-0.50$ V up to $-2.00$ V vs SHE as reducing potentials.

\begin{figure*}[h!]
\centering
\includegraphics[width=0.9\textwidth]{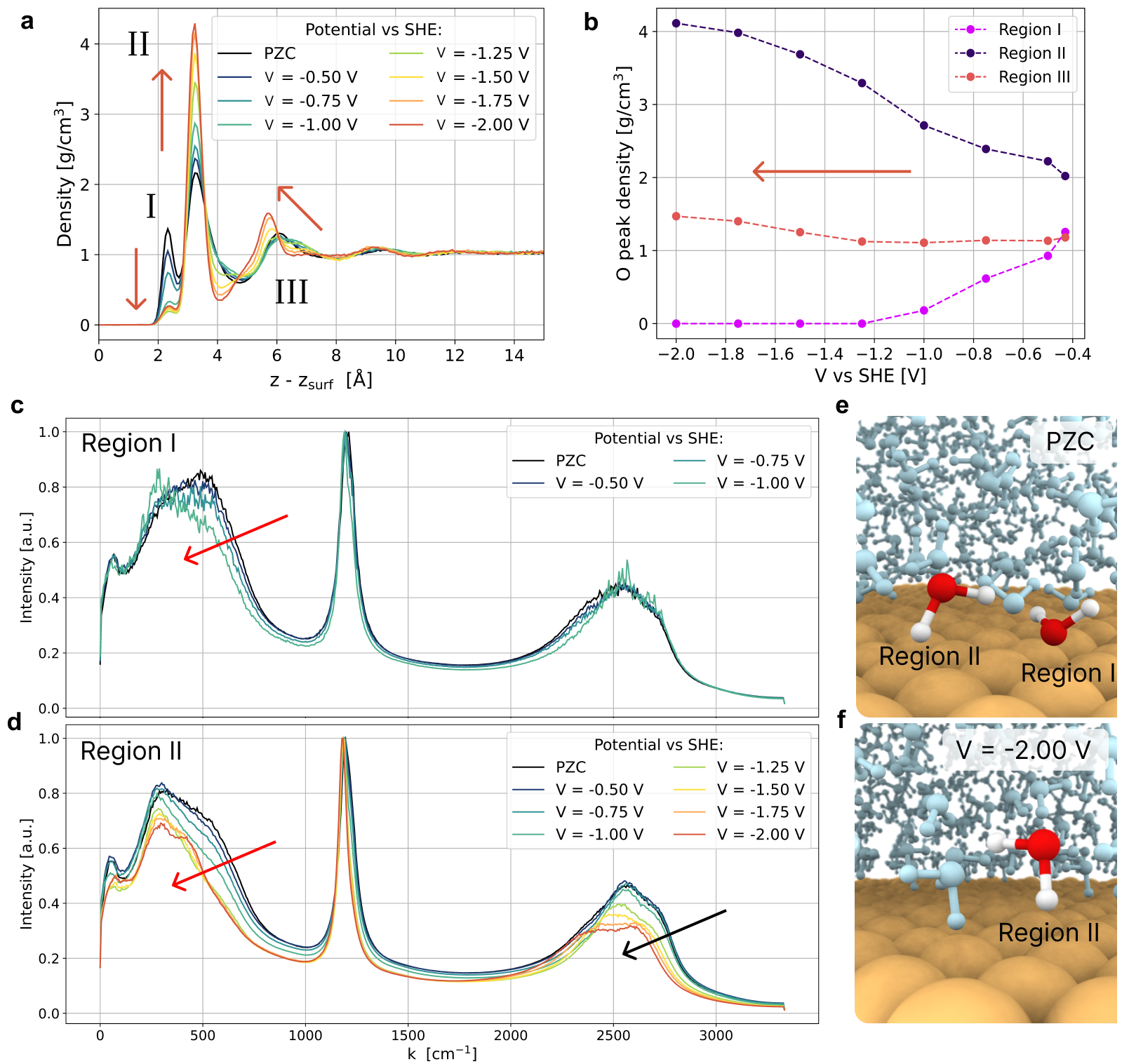}
\caption{Properties of interfacial water. (a) Density profile in the solvent region vs the distance z with respect to the surface z\textsubscript{surf}, at different values of external potential. Arrows emphasise trends moving towards more negative potentials. (b) Summary of the oxygen peak density height in the interfacial regions vs the applied potential. VDOS spectra for hydrogen atoms in region I (c) and region II (d). From low to high frequencies, it is possible to identify the peaks associated with the H-bond bending ($\approx$ 40~$\text{--}$~50~cm\textsuperscript{-1}), the libration (150~$\text{--}$~700~cm\textsuperscript{-1}), the H-O-H bending ($\approx$~1200~cm\textsuperscript{-1}), the O-H symmetric and asymmetric stretching (2300~$\text{--}$~2800~cm\textsuperscript{-1})~\cite{le_theoretical_2018,jin_temperature_2024}. Spectra are computed using the deuterium mass for the hydrogen species. Frames of the MD trajectory in which specific molecules are emphasised to show the typical water orientation in regions I and II at PZC (e) and at $-2.00$ V vs SHE (f). O and H atoms of emphasised molecules are depicted in red and white, respectively.}
\label{fig:application}
\end{figure*}
\subsub{Structure of interfacial water}
Firstly, we examine the density profile of interfacial water at different potentials (Fig.~\ref{fig:application}a). Looking at these profiles, it is possible to identify, at least, three visible regions in the solvent, close to the interface. These regions are sequentially labelled as I, II and III, moving from the surface towards the water bulk. There are still some minimal fluctuations in the density after 8~\text{\AA} far from the surface, but they are completely smoothed out at 12~\text{\AA}.\\
The most evident signature of the applied potential is the variation in the magnitude of the peak density (Fig.~\ref{fig:application}a). The density of peak I drastically decreases at increasing negative potentials, and it almost disappears for \textsf{V}~$\leq$~$-1.25$~V. A more detailed analysis, focusing only on the oxygen role (Fig.~\ref{fig:application}b), reveals that the O contribution in region I is completely suppressed for \textsf{V}~$\leq$~$-1.25$~V: there are no more \water molecules in region I, and the residual contribution to the total density in region I is associated with hydrogens of region II molecules.\\
The application of the bias also causes the density in peak II to increase to a value 4.5 times larger than the bulk water one (Fig.~\ref{fig:application}b). As discussed in the Supplementary Section~\ref{SI:dipole_analysis}, this peak modulation is associated with a variation in the preferential orientation of the water dipoles. Region I is composed of molecules with oxygen atoms exposed towards the surface and hydrogen atoms pointing away towards other water molecules (Fig.~\ref{fig:application}e). This orientation is compatible with a chemisorption process due to a dative bond between the oxygen lone pair and a copper atom. On the contrary, molecules in region II are characterised by an orientation in which H atoms are mainly exposed towards the surface (Fig.~\ref{fig:application}e): these molecules are considered as physisorbed. This dipole flip mechanism with increasing negative bias has already been reported~\cite{toney_distribution_1995,raffone_revealing_2025} and is related to the variation of the strength of the Cu$\cdot \cdot \cdot$O  chemical interaction and the Cu$\cdot \cdot \cdot$H  electrostatic attraction at variable surface charge. At reducing potentials, the increase of the negative surface charge enhances the repulsion of negative oxygen atoms while attracting the positive hydrogens, flipping the water molecules (Fig.~\ref{fig:application}f). These effects are also illustrated in the Supplementary Video through MD snapshots, highlighting the Bader charge computed on the electrode at the DFT level and the behaviour of interfacial molecules.\\
The effectiveness of \cosot in studying multiple PESs is demonstrated by its ability to replicate these well-known effects and, more importantly, to produce monotonous and consistent trends among the various potential values.
Even if AIMD can already give some information about these phenomena, our model provides a more detailed picture, affording an analysis at several applied potentials and without finite-size effects. In particular, two new elements are identified: an evident transition in region I-II between $-1.00$ and $-1.25$ V and a more complex behaviour of region III. As soon as peak I disappears, between $-1.00$ and $-1.25$ V, there is an abrupt increase in the O density in region II, sharpening the peak. At the same time, this transition also has consequences in region III. Here, the O density is almost unperturbed at low voltages (Fig.~\ref{fig:application}b), but peak III markedly increases at higher negative values. The bias influences not only the height of the peak but also its position: peak III tends to shift closer to the surface at more negative \textsf{V} ($\Delta$z\textsubscript{peak}~$\approx$~0.3~\text{\AA} at \textsf{V}~$=$~$-2.00$~V vs SHE). To the best of our knowledge, the effects of the applied potential in region III have never been discussed before. Indeed, this region and the complete transition to the water bulk are commonly not accessible via AIMD due to the computational cost of considering water layers thicker than 6~$\text{--}$~8 \text{\AA}. 

\subsub{Features of the interfacial H-bond network}
Another important feature of interfacial water is the strength of the H-bonds and how this is altered by the applied bias. This information can be inferred from the analysis of the vibrational density of states (VDOS) spectra for H atoms (Fig.~\ref{fig:application}c-d). For example, the relative strength of the H-bonds vs the O-H covalent bonds in water is correlated to O-H stretching vibration modes~\cite{le_theoretical_2018}. As emphasised with the black arrow in Fig.~\ref{fig:application}d, the O-H stretching band in region II visibly red-shifts. This shift is the typical signature of the weakening of the O-H covalent bond due to stronger H-bonds~\cite{le_theoretical_2018}.\\ 
The stretching band reflects the strength of H-bonds through the relative motion of oxygen and hydrogen atoms along the covalent bond axis, while the libration modes capture the restricted rotational motion of water molecules caused by H-bonding interactions~\cite{tong_experimentally_2016}. As emphasised by the red arrows in Fig.~\ref{fig:application}c-d, the libration band in regions I and II is altered by the applied bias. The shift at lower energies in region I is proof that molecules can more easily rotate in region I due to the weaker Cu$\cdot \cdot \cdot$O  interaction. Whereas, in region II, the high-energy shoulder of the libration band is more suppressed at increasing negative potentials. This red-shift can be interpreted as the signature of a less restricted rotational motion at increasing negative potentials due to a smaller number of H-bonds in region II (H atoms pointing towards the surface are not involved in H-bonds with other molecules).\\
It is important to note that computing VDOS spectra via AIMD is often limited by short simulation times, which hinder statistical convergence and hide spectral features. In contrast, the spectra presented here benefit from nanosecond-scale sampling, significantly improving the resolution of peak positions and revealing fine details in the spectra. This enhanced statistical quality allows for a more reliable interpretation of bias-induced changes in interfacial water dynamics.

\section{Discussion}
We presented \coso, a general and data-efficient workflow for developing ML-FFs to investigate metal-\water interfaces under external electrical bias. This enables a high-fidelity description of these systems by combining a fully explicit solvent model, a DFT-accurate representation of the liquid water network via the SCAN functional, and a constant potential framework that avoids explicit-implicit solvent compromises through the ``double reference method''.\\
To address the computational cost of this rigorous setup, \cosot leverages a data-efficient transfer learning algorithm for the ML-FFs training and an effective active learning protocol. To ensure model stability in the early stages of the ML-FF development, we perform transfer learning from general-purpose potentials. In subsequent iterations, we refine the ML-FFs by transferring information from domain-specific models to achieve a superior accuracy. This dual strategy enables the construction of robust and precise ML-FFs across a wide range of potential values. Our workflow achieves stable MD trajectories with only a few hundred configurations, reaches the same accuracy as models trained from scratch, and converges physical observables with a final dataset of just one thousand configurations. This data efficiency is critical for enabling DFT-level labelling in constant potential simulations, especially with our computationally demanding settings.
Importantly, the applicability of the \cosot workflow is not constrained by our specific \textit{ab initio} setup used for data labelling. Its modular design allows for easy customisation (in the choice of DFT functional, electrification scheme, or pre-trained model for transfer learning), making it adaptable to the requirements of diverse electrochemical interfaces, not limiting to metal-\water systems.\\
To demonstrate its capabilities, we applied the \cosot workflow to Cu-water interfaces as a representative system. Our ML models successfully revealed detailed structural and dynamical properties of electrified metal-water interfaces, achieving a level of resolution rarely accessible with conventional approaches. This picture can be further enriched by incorporating surface defects, ions, adsorbates, and electrocatalytic reactions to simulate specific \textit{operando} conditions.
While these more complex scenarios introduce new challenges, such as accounting for long-range electrostatic interactions from charged species, the core strengths of \cosot remain unaffected. Its data-efficient and robust protocol provides a solid foundation for developing more advanced ML-FFs in which the \cosot models are a modular element within a more complex architecture.
For these reasons, we believe that the \cosot strategy paves the way for high-quality, atomistic simulations of electrified interfaces, driving progress in both the development of modelling techniques and the advancement of fundamental and applied research in electrochemistry.

\section{Methods}\label{method}

\subsection{Details of DFT calculations}
\label{DFT_method}
The \textit{ab initio} calculations required to generate data for the ML dataset are based on non-polarised DFT simulations, as implemented in the \textit{Vienna Ab initio Simulation Package} (\texttt{VASP}), version 6.4.3~\cite{Vasp1,Vasp2,Vasp3}.
The projector-augmented-wave (PAW) method is employed to describe the electron-ion interaction~\cite{PAW_VASP}. Electronic wave functions are expanded in plane waves with a cut-off energy of 800 eV in conjunction with a dense FFT grid to properly guarantee numerical convergence of energies as well as of forces. The Brillouin zone is sampled by employing a Gamma-centred (11 11 1) Monkhorst-Pack mesh in the case of the unit cell of the Cu slab and consistently reduced for the supercells. A Gaussian smearing with a spreading of 0.1 eV is employed. Gamma-point-only calculations are performed for systems including only \water molecules.\\
The meta-generalised-gradient approximation (meta-GGA) ``Strongly Constrained and Appropriately Normed'' (SCAN) functional is employed~\cite{Chen_PNAS_2017,sun_PRL_2015}. Indeed, it is well known that common GGA functionals, such as PBE, tend to markedly over-structure the liquid water network~\cite{Chen_PNAS_2017,lacount_ensemble_2019}. Whereas, SCAN succeeds in describing the \water behaviour due to its capability to reproduce the correct relative magnitude of the covalent O-H bonds, the hydrogen bonds and the dispersion forces that determine the H-bond network~\cite{Chen_PNAS_2017}.  
In view of the numerical instability and convergence issues associated with the original formulation of the SCAN functional, the revisited r\textsuperscript{2}SCAN version is adopted in this work~\cite{r2scan_2020}.

\subsection{Interface electrification scheme}\label{double_ref}
The electrification of the interface at the DFT level is performed following the ``double reference method''~\cite{taylor_first_2006}.
By allowing the surface charge to adjust to reach a target potential and compensating with a homogeneous background charge for the added or removed electrons, this approach is compatible with an electronically grand-canonical ensemble~\cite{hagopian_advancement_2022}. The robust definition of the electrode potential is established via a chain of potential references involving two auxiliary systems such that this definition is not influenced by the counter-charge and any fictitious solvent-vacuum interface. Specifically, the first auxiliary system has no extra charge (i.e., no applied potential) and a vacuum region to define the vacuum level; the second is with no extra charge and no vacuum and represents a reference for the same system but with extra charge (i.e., with applied potential).\\ 
Here, we employ the definition of the electrode potential within the ``double reference method'' to accomplish two tasks. First, by applying this scheme to the configurations in the PZC dataset, we estimate the potential of zero charge to be approximately $-0.1$ to $-0.2$ V vs SHE. This value serves as an upper bound for the range of reducing potentials. Second, by using the same method on geometries with added or removed charge, we establish a direct link between the net charge and the corresponding applied bias. This enables us to define a criterion for tuning the system charge to reach a target bias value. In this approach, the same geometry is labelled at multiple bias values. These calculations share the chain of references. As a result, the computational cost of the auxiliary systems is distributed across datasets corresponding to different applied biases.\\
Practically, the grand-canonical DFT is performed by modifying the ``Fully Constant Potential'' (FCP) \texttt{ASE} calculator proposed in Ref.~\cite{xia_grand_2023}, with the definition of the electrode potential in the ``double reference method''. This calculator is then integrated into a Python routine that implements an automatic workflow to evaluate the auxiliary systems of the ``double reference method'' and perform constant potential calculations at fixed geometry for more target potential values (see Code Availability).  

\subsection{Machine learning force-fields}\label{section_ML}

\subsub{Constant potential models via Franken transfer learning}
The training of the constant potential force-fields is performed within the \Frankent transfer learning approach~\cite{novelli_fast_2025}. This scheme is based on the integration of the accurate descriptors of a pre-trained deep neural network with an efficient large-scale kernel regression approach. The kernel function is approximated by means of Random Fourier Features (RF) maps to overcome the scalability issues of kernel methods in the case of large datasets~\cite{dhaliwal_machine_2022}. This structure allows for fast training and inference typical of kernel methods, but irrespective of the dataset size.\\
As shown in Fig.~\ref{fig:scheme_transfer_learning}, starting from common descriptors (the backbone) extracted as the node features of a graph neural network (GNN), different readout blocks are used to target specific bias values: we trained seven RF-based readouts for potentials ranging from $-0.50$ V to $-2.00$ V vs SHE (with a spacing of 0.25 V). This approach is inherently scalable, enabling the simultaneous generation of a greater number of regression blocks that encompass an even broader spectrum of applied biases. 
In addition to using descriptors from the pre-trained general-purpose potential \texttt{MACE-MP0}, as done in Ref.~\cite{novelli_fast_2025}, we employ domain-specific ones. That is, we extract the descriptors from a MACE model trained on a set of data of the same interface but without an applied electric field (cf. Section ``\MACEt model for the PZC'').  While this approach requires training an additional model, the computational cost of labelling configurations at PZC is significantly lower (approximately 10 to 20 times less than performing full constant potential calculations across seven target bias values). Therefore, the increased accuracy with a reduced number of grand-canonical DFT evaluations justifies the additional effort required to develop the PZC model.\\
After testing different descriptors and numbers of RFs (Supplementary Section~\ref{SI:test_RF}), final results are achieved with the descriptors from our PZC \MACEt model and 8192 RFs.  

\subsub{MACE model for the PZC}
The ML force-fields (ML-FFs) at the PZC are built using the \MACEt software, version 0.3~\cite{batatia_arxiv_2022}. The peculiarity of \MACEt is the construction of a high-order message-passing scheme based on a many-body atomic cluster expansion (ACE) representation and the adoption of equivariant internal features for the message~\cite{batatia_arxiv_2021}. It results in a data-efficient structure with appreciable in-domain and out-domain predictions.
Specifically, we employ a 4-body term expansion and two interaction layers, each with a cut-off radius of 6 \text{\AA}. Considering the possible presence of non-local contributions outside this receptive field, we test models with cut-off radii up to 9 \text{\AA}, as well as a model with a cut-off of 5 \text{\AA} but with three interaction layers (Supplementary Section~\ref{SI_cutoff}). In all cases, we observe that the RMSE of the force predictions decreases only marginally as the receptive field increases. At the same time, the solvent density profile remains unchanged with the expansion of the receptive field, indicating that non-local contributions do not significantly affect the solvent structure. These results demonstrate that the selected cut-off radius and number of interaction layers are sufficient to capture all relevant interactions within the system.\\
During the construction of the ML-FF, an architecture with equivariant messages ($L = 1$) and 128 channels is employed to better optimise the stability of the model in case of a limited-size database. Once the dataset is complete, the final ML-FF is trained with invariant messages ($L = 0$) and 256 channels, resulting in a reasonable balance between accuracy and computational efficiency.
The dataset is split into training/validation/test subsets with a ratio of 85:10:5. The model is optimised with the AMSGrad algorithm, using a learning rate of 0.01, a batch size of 4 and a maximum number of epochs equal to 1000. Performance is evaluated on energy and forces with a weighted root mean square error (RMSE) loss function. The weights for energy and forces in the loss function are initially set to 1 and 100, respectively. In the last 20\% of the training, the weight for energy is increased to 1000.

\subsub{Dataset generation}\label{data_gen}
Different strategies are adopted to generate a preliminary dataset for the PZC. A pre-existing dataset~\cite{H2O_dataset} is used to describe the pure water bulk. Configurations for bare Cu surfaces are included after sampling AIMD trajectories at a lower accuracy level (i.e., PBE). A ML universal model~\cite{batatia_foundation_2024} is also used to collect data for Cu-\water systems. For all these cases, the configurations are labelled consistently with our DFT setup.\\
Starting from this initial pool of configurations, the training dataset is iteratively refined through the ``Data-Efficient Active Learning'' (\DEAL) scheme, recently introduced in Ref.~\cite{perego_data_2024}. This method allows us to progressively enlarge the dataset and refine the ML-FF in a data-efficient manner, providing new configurations selected from MD trajectories obtained with preliminary ML-FFs. Firstly, the MD configurations are screened, relying on the uncertainty prediction obtained by a \textit{query-by-committee} approach. To avoid redundant labelling, this set of selected configurations is then filtered by \DEALt using a sparse Gaussian process model, trained on-the-fly on the pre-selected data. More details on this approach are reported in the original article~\cite{perego_data_2024}. This strategy ensures broad and uniform coverage of the relevant configuration space while dramatically reducing the number of costly \textit{ab initio} calculations.\\
At the end of the active learning, the complete dataset for the Cu-\water interface at the PZC is composed of about 5320 configurations, including systems of bulk \water, bare Cu surfaces and Cu-\water interfaces.\\
A similar iterative strategy is used to construct datasets at multiple applied potentials. Starting from a subset of 250 configurations sampled from the PZC dataset, models are progressively refined using the \DEALt protocol. In this case, a two-step \DEALt selection is performed. First, configurations are filtered independently for each applied bias; then, the selected configurations are merged across all bias values and screened again alongside the existing dataset. This second \DEALt selection reduces redundancy among configurations sampled at the different biases and with respect to the existing dataset.
This iterative refinement continues until the \cosot models reach an accuracy comparable to the reference PZC model and demonstrate stable convergence of key physical observables.

\subsection{Molecular dynamics simulations}
\label{MD_method}
Classical molecular dynamics (MD) simulations are performed with the \textit{Large-scale Atomic / Molecular Massively Parallel Simulator} (\texttt{LAMMPS}) software~\cite{LAMMPS}, supplemented by \MACEt v0.3~\cite{batatia_arxiv_2022}. NVT simulations are performed with an integration step of 0.5 fs, and the deuterium mass is used for the hydrogen atoms to properly describe the dynamics of light atoms. The temperature is controlled using a Nosé–Hoover thermostat chain with a damping parameter of 100 times the integration step. The thermostat temperature is properly set to ensure that the outermost metal layers of the slab and water molecules are in thermodynamic equilibrium at 330 K. This slight increase above room temperature is required to overcome the residual over-structuring of water predicted by SCAN~\cite{Chen_PNAS_2017}. The dynamical properties are computed in the microcanonical ensemble to remove any spurious contribution from the thermostat. For the VDOS spectra, the velocity autocorrelation function is evaluated by recording the velocities every 5 fs for several time windows of 6 ps. 
MD runs are performed in a slab system composed of six layers of metal atoms, with the two innermost ones fixed. Periodic boundary conditions are imposed in all directions, and no regions of vacuum are present.
MD runs are performed in a slab system with a lateral size of 10~$\text{--}$~15 \text{\AA} (4$\times$4 up to 6$\times$6 supercells) during the active learning phase and 20 \text{\AA} (8$\times$8 supercells) in the final production run. In the latter case, a region of 30 \text{\AA} (i.e., 15 \text{\AA} per side) is filled with water molecules.

\bmhead{Supplementary information}
Supplementary information: Supplementary Sections 1–3, Figures 1–4.\\
Supplementary video: Illustrative video on the effects of the applied potential on the surface charge and the solvent structure. This material is available at~\url{https://bit.ly/copper_water_bias_movie}.

\bmhead{Acknowledgements}
The authors acknowledge the support of the Data Science and Computation Facility at the Fondazione Istituto Italiano di Tecnologia and the CINECA award under the ISCRA initiative. M.G.B. is grateful to F. Raffone, P. J. Buigues, and S. Perego for valuable discussions and to P. Novelli and G. Meanti for providing the Franken code. L.B. and M.P. acknowledge funding from the European Union - NextGenerationEU initiative and the Italian National Recovery and Resilience Plan (PNRR) from the Ministry of University and Research (MUR), under Project PE0000013 CUP J53C22003010006 ``Future Artificial Intelligence Research (FAIR)''. The authors are also grateful for the financial support from the National Recovery and Resilience Plan (NRRP), Mission 4 ``Education and Research'' – Component 2 ``From research to business'' - Investment 3.1 ``Fund for the realization of an integrated system of research and innovation infrastructures'' – Call n. 3264 of 28/12/2021 of Italian Ministry of University and Research | Award Decree n. 128 (21/06/2022)- Project code: IR0000027, Concession Decree No. 128 of 21/06/2022 adopted by the Italian Ministry of Research, CUP: B33C22000710006, Project title: iENTRANCE. This work also received support by the European Innovation Council (EIC) under the European Union’s Horizon Europe research and innovation programme through the project PREDICT (Grant Agreement No. 101219185). The views and opinions expressed are those of the authors and do not necessarily reflect those of the European Union or the European Innovation Council. Neither the European Union nor the granting authority can be held responsible for them. Finally, this publication is supported by the COST Action CA24103: EUropean COllaborative Network on electro-Catalysis for Efficient Renewable Technologies (EU-CONCERT) and the flagship program ``Teaching Science to Computers'' of the Italian Institute of Technology.

\section*{Declarations}

\bmhead{Data availability}
A minimal dataset to evaluate the \cosot workflow is provided alongside the code at~\url{https://github.com/michelegiovannibianchi/TRECI}. The molecular dynamics trajectories underlying the analysis of the interface
 are available on Zenodo at~\cite{zenodo_repo}. 

\bmhead{Code availability}
The \cosot code underlying this work is freely available at~\url{https://github.com/michelegiovannibianchi/TRECI}. The FCP calculator employed in \cosot is available at~\url{https://github.com/michelegiovannibianchi/DoubleReferenceMethod-FCP-calculator}. \cosot workflow requires the \MACEt code available at~\url{https://github.com/ACEsuit/mace}, the \Frankent code available at~\url{https://github.com/CSML-IIT-UCL/franken} and the \DEALt code available at~\url{https://github.com/luigibonati/DEAL}.

\bmhead{Author contribution}
L.B. and G.C. designed and supervised the experiment. M.G.B. implemented the code and performed the calculations. All authors analysed the simulations. M.G.B. and L.B. wrote the first draft, and all authors revised it.

\bmhead{Competing interests}
The authors declare no competing interests.

\bibliography{sn-bibliography}

\appendix
\clearpage
\onecolumn
\numbered

\renewcommand{\figurename}{Supplementary Figure}
\renewcommand{\thefigure}{S\arabic{figure}}
\renewcommand{\tablename}{Supplementary Table}
\renewcommand{\thetable}{S\arabic{table}}
\renewcommand{\thesection}{S\arabic{section}}
\renewcommand{\thepage}{S\arabic{page}}

\renewcommand{\theHfigure}{S\arabic{figure}}
\renewcommand{\theHtable}{S\arabic{table}}
\setcounter{page}{1}
\setcounter{section}{0}
\setcounter{figure}{0}
\setcounter{table}{0}

\begin{center}
{\Large{Supplementary Information}}\\
\bigskip

\Large{\textbf{Electrified Metal-Water Interfaces at Constant Potential:\\ Data-Efficient Transfer Learning for\\ Machine-Learning-Based Molecular Dynamics}}

\bigskip 
{\large{
Michele Giovanni Bianchi$^{1,2}$, Michele Re Fiorentin$^{1}$, Francesca Risplendi$^{1}$, Candido Fabrizio Pirri$^{1,3}$, Michele Parrinello$^{2}$, Luigi~Bonati$^{2,*}$, Giancarlo Cicero$^{1,*}$}\\
}
\vspace{.4cm}
{\large{
\textit{$^1$ Department of Applied Science and Technology, Politecnico di Torino, corso Duca degli Abruzzi 24, Torino, 10129, Italy}\\
\textit{$^2$ Atomistic Simulations, Italian Institute of Technology, via Enrico Melen 83, Genova, 16152, Italy}\\
\textit{$^3$ Centre for Sustainable Future Technologies, Italian Institute of Technology, via Livorno 60, Torino, 10144, Italy}

\vspace{.4cm}
*Corresponding authors. E-mails: luigi.bonati@iit.it; giancarlo.cicero@polito.it}}
\end{center}

\section{Additional tests on the constant potential models}

\subsection{Accuracy test on the descriptors and the number of random features}\label{SI:test_RF}
\begin{figure*}[h]
    \centering
    \includegraphics[width=0.75\linewidth]{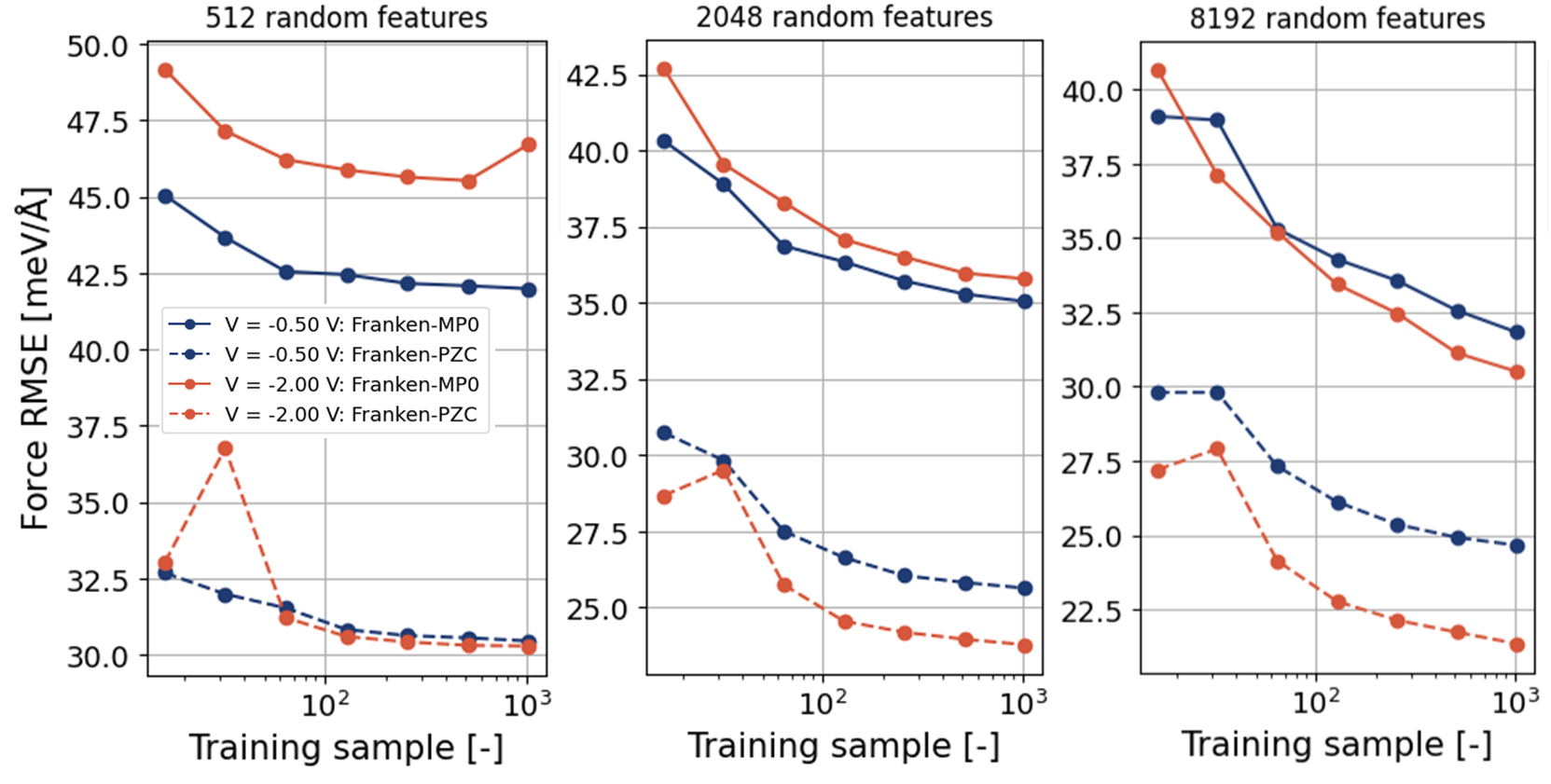}
    \caption{Comparison of the RMSE error on force prediction at \textsf{V} = -0.50 V and -2.00 vs SHE, using different descriptors (\FrankenMPt vs \FrankenPZC) for different training set sizes and numbers of random features.}
    \label{SI:fig_backbone_RF}
\end{figure*}

Figure~\ref{SI:fig_backbone_RF} shows a comparison of the accuracy of the \Frankent models based on \texttt{MACE-MP0} (i.e., \FrankenMP) and on the domain-specific PZC descriptors (i.e., \FrankenPZC).
Accuracy is evaluated on the force RMSE on a validation set of about 150 configurations. This analysis is performed by varying the size of the training set (from 16 up to 1024 configurations) and the number of the random features used to approximate the kernel function. We evaluate the cases where \textsf{V} is -0.50 and -2.00 V vs SHE, as these represent the nearest and most distant conditions for the descriptors trained on the PZC geometries.
It is evident that the domain-specific PZC descriptors are superior to the general-purpose \texttt{MACE-MP0} ones at -0.50 V as well as at -2.00 V, regardless of the training set size and the number of random features. 
This test further proves the importance of the domain-specific descriptors in the \cosot strategy. 

\subsection{Assessing the stability of domain-specific models}\label{SI_Extra_backbone}
As discussed in the main text, models based on domain-specific descriptors generally require more data than general-purpose models to ensure stable MD trajectories, particularly in cases involving highly negative potentials.
To address this, we tested \Frankent models built from various domain-specific models, including both equivariant GNNs and larger datasets.\\
As expected, models employing equivariant descriptors proved more expressive, delivering stable results with smaller datasets compared to their invariant counterparts. This is the case of the descriptors from a \MACEt model with equivariant messages ($L = 1$) and 128 channels, trained exclusively on geometries at PZC (\texttt{Franken-PZC-L1}). However, equivariant architectures are computationally more demanding, increasing inference costs.\\
For this reason, we also explored invariant descriptors trained on larger datasets. One such case is a \MACEt model with invariant messages ($L = 0$) and 256 channels, trained on a combined dataset comprising PZC geometries and configurations compatible with applied potentials (\texttt{Franken-PZC+Charged-L0})\footnote{Geometries compatible with applied potentials were relabelled without extra charge to maintain consistency with the PZC dataset. This relabelling incurs no additional cost, as this DFT calculation is already required for the auxiliary systems in the ``double reference method''.}. Within this framework, the active learning process enables not only the progressive optimisation of the readout blocks but also the retraining of descriptors using the expanded dataset, which includes the newly acquired geometries associated with the applied potential from the previous iterations.
Interestingly, after completing the active learning, we observed on the final dataset that even domain-specific MACE descriptors with invariant messages ($L = 0$) and 256 channels, trained only on PZC geometries (\texttt{Franken-PZC-L0}), can guarantee stable results. This demonstrates that stable performance can be achieved with a thousand of data points without resorting to expensive equivariant models or continuously retraining the descriptors.\\
Based on these findings, we recommend using equivariant descriptors for the dataset expansion during the active learning process. Whereas, for production MD runs, invariant descriptors are preferable due to their lower computational cost during inference.

\subsection{Cross-check among different descriptors}\label{SI_different_backbones}
\begin{figure*}[h]
\centering
\includegraphics[width=\textwidth]{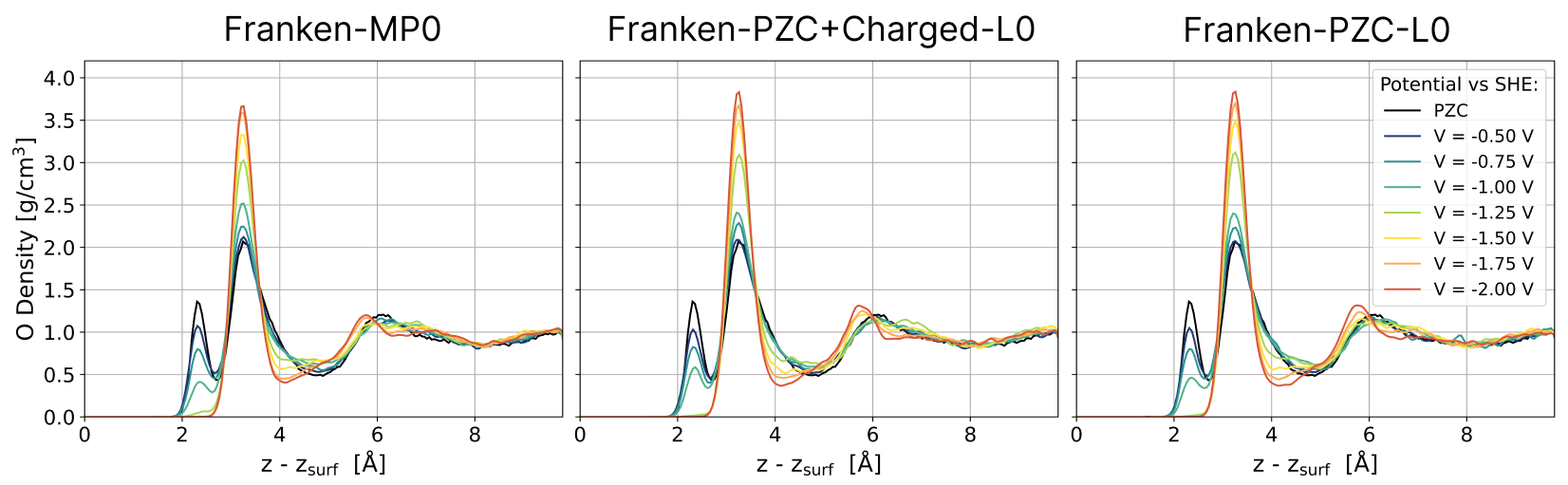}
\caption{Comparison of the oxygen density profiles of interfacial solvent at different applied potential values, as predicted by \Frankent models with different descriptors. The same PZC profile computed with a standard \MACEt model is reported in each panel for reference.}
\label{SI:fig_Odensity_backbone}
\end{figure*}

A characteristic of \cosot workflow is the monitoring of physical observables during the active learning to control their convergence and identify a condition to stop the iterative refinement of the model.
In addition, at the end of the active learning, we further validate the results: we compare a physical quantity (e.g., the solvent density profile) using models with different descriptors. As shown in Fig.~\ref{SI:fig_Odensity_backbone}, it is evident that the final results are almost independent from the employed descriptors, confirming that the generated small datasets are sufficient for accurate and reliable models.

\section{Test on the locality of the ML model}
An important parameter of an ML-FF is the receptive field, i.e., the spatial region around an atom that is considered in the determination of its interactions with neighbouring atoms. \MACE, and consequently \Frankent models based on \MACEt descriptors are able to expand the receptive field beyond the cut-off radius $r$ through a sequence of \textit{n} interaction blocks. The resulting nominal receptive field is $n \times r$. Here, we benchmark the effects of different cut-off radii and/or numbers of interaction blocks on the accuracy and on the predicted physical observables. Specifically, we perform these tests on the ML-FF for the PZC and for \textsf{V}~=~-~2.00 V.
\label{SI_cutoff}
\begin{figure}[h]
    \centering
    \includegraphics[width=\linewidth]{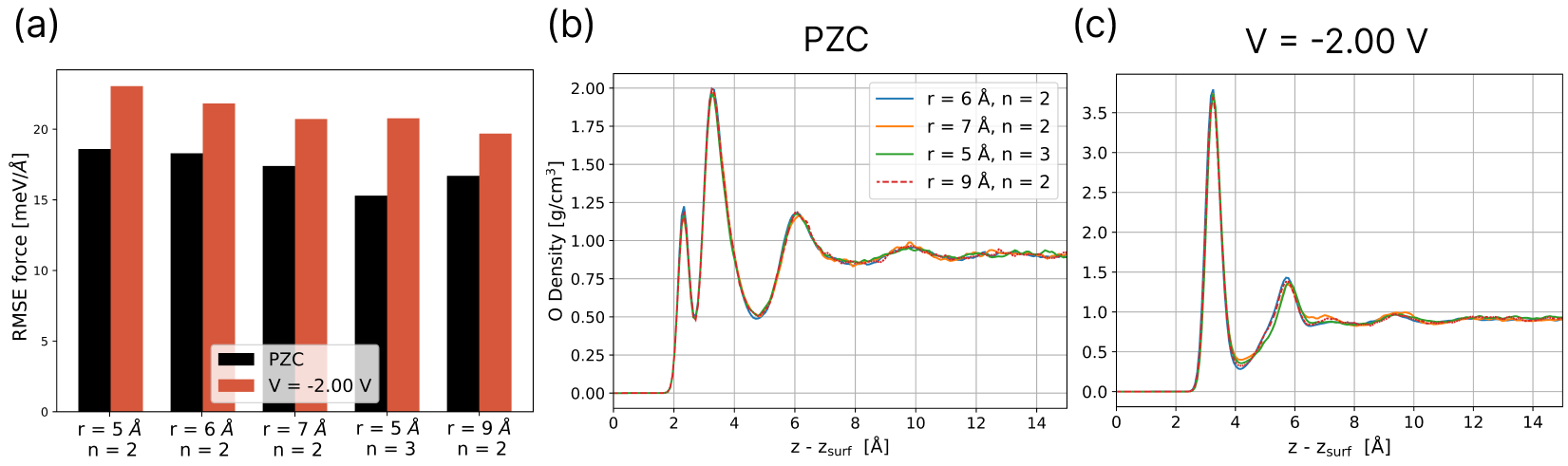}
    \caption{Comparison of the effects of different cut-off radii $r$ and/or number $n$ of interaction blocks on the accuracy of the models (a) and on the oxygen density profile (b-c) at the PZC and at \textsf{V}~=~-~2.00 V.}
    \label{SI:fig_cutoff}
\end{figure}
As shown in Fig.~\ref{SI:fig_cutoff}a, there is a minimal improvement in the RMSE on the force prediction as the receptive field is expanded for the PZC as well as for \textsf{V}~=~-~2.00 V.
We also evaluate if this small reduction of the error has an evident impact on the physical observables, such as the oxygen density profile.  Fig.~\ref{SI:fig_cutoff}b-c shows the O density profile at the PZC and at \textsf{V}~=~-~2.00 V for ML-FFs with different receptive fields. It is evident that the solvent density profile is independent from the receptive field of the ML model since a constant density is recovered within 12 \text{\AA} in all the cases. This proves that the recovery of the water bulk properties far from the surface is associated with the screening due to the water molecules rather than the local nature of the ML-FFs. For that reason, all the models are developed with $r$ = 6 \text{\AA} and $n$ = 2. 

\section{Additional analysis of the effects of the applied bias}
\label{SI:dipole_analysis}
\begin{figure*}[h]
    \centering
\includegraphics[width=\linewidth]{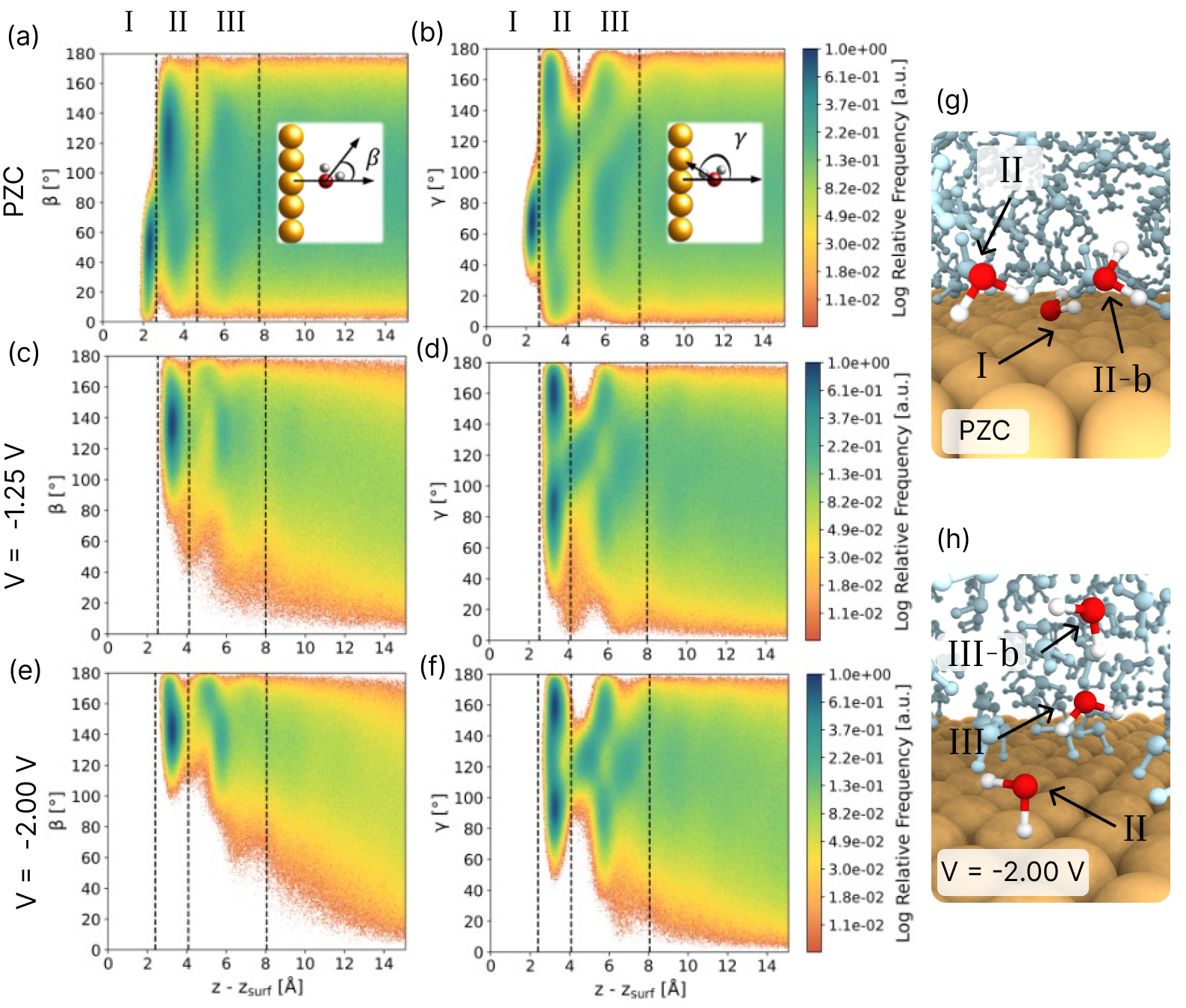}
    \caption{Colour maps for the distributions of the $\beta$ dipole and $\gamma$ O-H-bond angle at PZC (a-b), -1.25 (c-d) and -2.00 (e-f) V vs SHE. The vertical dashed lines identify the different regions. Frames of the MD trajectory in which specific molecules are emphasised to show the water orientation: (g) configuration of molecules in region I, region II and between region II and III (``Region II-b''), at low potentials, (h) configuration in region II, in region III close to region II (``Region III'') and in the middle of region III (``Region III-b'') at high negative potentials. O and H atoms of emphasised molecules are depicted in red and white, respectively.}
    \label{SI:fig_dipole}
    \end{figure*}

The modulation of the solvent density profile due to the applied potential, as discussed in the main text, is strictly correlated with a re-orientation of the dipole of water molecules. In this section, we more deeply analyse the water molecule orientation at the different potentials, considering the distribution of two angles. The first, $\beta$, is associated with the orientation of the dipole and is defined as the angle between the dipole vector of the molecule and the vector perpendicular to the surface (see inset of Fig.~\ref{SI:fig_dipole}a). Whereas the second angle, $\gamma$, is linked to the orientation of the O-H covalent bond and is defined as the angle between the vector of the O-H bond and the vector perpendicular to the surface (see inset of Fig.~\ref{SI:fig_dipole}b).\\
At the PZC (Fig.~\ref{SI:fig_dipole}a) and at low potentials (\textsf{V} $\geq$ -1.0 V vs SHE), region I (i.e., chemisorbed molecules) is characterised by a $\beta$ distribution peaked at around 40°~$\text{--}$~60°: the resulting molecular orientation has oxygen atoms exposed towards the surface and hydrogen atoms pointing away towards other water molecules, as visible in the MD frame reported in Fig.~\ref{SI:fig_dipole}g. Both hydrogens within the same molecule behave in a similar way and are at the same height with respect to the surface. This can be deduced from the distribution of the $\gamma$ angle in region I (Fig.~\ref{SI:fig_dipole}b): even if it is possible to define two $\gamma$ angles for each molecule, both distributions are equivalent and overlap in a unique unimodal distribution.\\ 
The picture is completely different in region II: here, the $\beta$ distribution reaches its maximum value in the range 120°~$\text{--}$~140°, meaning that the water dipole is mainly pointing towards the surface. This orientation is weakly dependent on the applied potential value: the peak of the $\beta$ distribution only slightly shifts at higher angles at more negative potentials due to a stronger electrostatic interaction between hydrogens and the surface atoms. The analysis of the $\gamma$ distribution provides additional details. For all the applied potentials, this distribution has the first lobe at about 140°~$\text{--}$~170°, where the O-H bond points towards the surface. The second lobe in the $\gamma$ distribution is at about 80°~$\text{--}$~110°. This is associated with the second O-H bond that is almost parallel to the surface, and its H atom is interacting with other water molecules within region II (see the orientations reported in Fig.~\ref{SI:fig_dipole}h).\\
In addition to the previously described general picture of water orientation, our ML-enhanced model provides a finer picture of the water structure. Indeed, a detailed analysis reveals a second weaker signal in the $\beta$ and $\gamma$ distributions from a sub-region between regions II and III: this is visible at PZC and progressively more blurred at increasing negative potentials. Looking at Fig.~\ref{SI:fig_dipole}a, it is possible to appreciate that there are a few molecules with a $\beta$ angle at about 40°~$\text{--}$~70° in region II. This contribution is associated with water molecules with dipoles pointing away from the surface, towards region III (see the  orientation reported as ``Region II-b'' in Fig.~\ref{SI:fig_dipole}g). This is also visible in the $\gamma$ distribution (Fig.~\ref{SI:fig_dipole}b) with a contribution at around 90°~$\text{--}$~110° (O-H bond parallel to the surface and H interacting with other molecules of region II) and at around 10°~$\text{--}$~30° (O-H bond pointing towards region III and H interacting with molecules of region III). Notice that this detail is usually not accessible in standard AIMD due to the high signal-to-noise ratio of the statistics but provides interesting information: molecules passing from region II to III (and vice versa) have to flip the dipole with a $\beta$ angle from 120°~$\text{--}$~140° to 40°~$\text{--}$~70°. This observation is valid at PZC and at low potentials (\textsf{V} $\geq$ -1.0 V).\\
The description completely changes at \textsf{V} = -1.25 V (Fig.~\ref{SI:fig_dipole}c-d) and at more negative potentials (Fig.~\ref{SI:fig_dipole}e-f). The applied potential has an impact on the structure of region III, where it becomes evident that there are two water orientations, at least. The dipole angle is about 160°~$\text{--}$~180°, while $\gamma$ is about 110°~$\text{--}$~130° in region III in proximity to the edge with region II. These contributions are associated with water molecules with the dipole perpendicular and pointing towards the surface. Both hydrogen atoms are almost at the same height (single spot for $\gamma$ at about 110°~$\text{--}$~130°) and are interacting with O of region II. This orientation can be visualised in the MD frame of Fig.~\ref{SI:fig_dipole}h, where this orientation is identified as ``Region III''.
On the contrary, the $\beta$ and $\gamma$ distributions in the middle of region III are more similar to the region II one. An O-H bond is perpendicular to the surface ($\gamma$ at 140°~$\text{--}$~170°) and the other one is almost parallel to it ($\gamma$ at 70°~$\text{--}$~100°), as emphasised in Fig.~\ref{SI:fig_dipole}h, with the label ``Region III-b''. The resulting dipole is not exactly perpendicular to the surface (differently from region III close to region II) but $\beta$ is peaked at 110°~$\text{--}$~150°.\\
We emphasise once more that these findings cannot be obtained through conventional AIMD; however, they could play a crucial role in surface reactivity. At considerably negative potentials, molecules that interact directly on the surface (region II) are not in contact with an area resembling bulk water; rather, they can migrate in a zone that retains a preferential molecular orientation (region III).

\end{document}